\documentclass[preprint,11pt]{aastex}

\newcommand{\simle}{\mbox{$\stackrel{<}{_{\sim}}$}}
\newcommand{\simge}{\mbox{$\stackrel{>}{_{\sim}}$}}

\slugcomment{To Appear in Astrophysical Journal}
\shorttitle{Near-IR Disk Sizes}
\shortauthors{Monnier \& Millan-Gabet}

\begin{document}

%% LaTeX will automatically break titles if they run longer than
%% one line. However, you may use \\ to force a line break if
%% you desire.
\title{On the Interferometric Sizes of Young Stellar Objects}

%% Use \author, \affil, and the \and command to format
%% author and affiliation information.
%% Note that \email has replaced the old \authoremail command
%% from AASTeX v4.0. You can use \email to mark an email address
%% anywhere in the paper, not just in the front matter.
%% As in the title, you can use \\ to force line breaks.

\author{J. D. Monnier %\altaffilmark{1} 
and R. Millan-Gabet %\altaffilmark{2}
}

\affil{Harvard-Smithsonian Center for Astrophysics,
60 Garden Street, Cambridge, MA, 02138}

%\altaffiltext{1}{University of Michigan}
%\altaffiltext{2}{Interferometry Science Center}

\email{jmonnier@cfa.harvard.edu, rmillan@cfa.harvard.edu}

%% Mark off your abstract in the ``abstract'' environment. In the manuscript
%% style, abstract will output a Received/Accepted line after the
%% title and affiliation information. No date will appear since the author
%% does not have this information. The dates will be filled in by the
%% editorial office after submission.

\begin{abstract}

Long-baseline optical interferometers can now detect and resolve hot
dust emission thought to arise at the inner edge of circumstellar
disks around young stellar objects (YSOs).  We argue that the
near-infrared sizes being measured are closely related to the radius
at which dust is sublimated by the stellar radiation field.  We
consider how realistic dust optical properties and gas opacity
dramatically affect the predicted location of this dust destruction
radius, an exercise routinely done in other contexts but so far
neglected in the analysis of near-infrared sizes of YSOs.  We also
present the accumulated literature of near-infrared YSO sizes in the
form of a ``size-luminosity diagram'' and compare with theoretical
expectations.  We find evidence that large ($\simge$1.0\,$\mu$m) dust
grains predominate in the inner disks of T~Tauri and Herbig Ae/Be
stars, under the assumption that the inner-most gaseous disks are
optically-thin at visible wavelengths.

\end{abstract}

%% Keywords should appear after the \end{abstract} command. The uncommented
%% example has been keyed in ApJ style. See the instructions to authors
%% for the journal to which you are submitting your paper to determine
%% what keyword punctuation is appropriate.

%% From the front matter, we move on to the body of the paper.
%% In the first two sections, notice the use of the natbib \citep
%% and \citet commands to identify citations.  The citations are
%% tied to the reference list via symbolic KEYs. The KEY corresponds
%% to the KEY in the \bibitem in the reference list below. We have
%% chosen the first three characters of the first author's name plus
%% the last two numeral of the year of publication as our KEY for
%% each reference.

%\tableofcontents

\section{Introduction}
Radically improved infrared (IR) detectors \citep{rmg1999b} have recently
allowed optical interferometers to investigate the inner accretion
disks around young stellar objects (YSOs) for the first time 
\citep{malbet1998,rmg1999,rmgphd,
akeson2000,rmg2001,lkha2001,mwc2001,akeson2002}.  In most cases,
the near-IR sizes were found to be significantly larger than
expected from the favored disk models of the time
\citep{malbet1991,hillenbrand1992,hartmann1993,cg97}.

These modeling efforts typically did not directly consider the near-IR
disk sizes; indeed, much of the model sophistication was directed
toward explaining the observed long-wavelength ($\lambda\simge10\mu$m)
spectral energy distributions.  Theoretical work has recently been
focused again on reproducing the near-IR properties of YSOs, both the
near-IR spectrum and characteristic sizes
\citep{natta2001,dullemond2001}.  The new models abandon the
optically-thick, spatially-thin gas disk in the inner-most region,
incorporating an {\em optically-thin} central cavity to explain the
``large'' near-IR sizes \citep[as suggested by][]{lkha2001}.
Alternatively, some workers argue for a quasi-spherical envelope of
dust grains around Herbig Ae/Be stars which also can explain the large
characteristic sizes \citep{mirosh1997,mirosh1999}.

This paper takes a step back from the increasingly complicated models
one encounters in the literature in order to develop a minimalist
framework useful for interpreting the new interferometric measurements
(e.g., as generic, or {\em model-independent}, as possible).  While
the essential physics are gradually being incorporated in
sophisticated, self-consistent (``physical'') models, theoretical
guidance is needed now in crafting and interpreting disk size surveys
beginning at the Keck Interferometer and the Very Large
Telescope Interferometer.  Indeed, most of the free parameters
in the physical models do not strongly affect the near-IR sizes
and only serve to obsfucate the relevant physics in the context of the
recent interferometric measurements.

In this Letter, we explore how basic dust grain properties and gas
absorption of the stellar ultraviolet continuum can affect the
observed characteristic sizes of YSO disks in the near-IR. 

\section{Compendium of Characteristic Disk Sizes}

\subsection{The Sample}
\begin{figure}[th]
\begin{center}
\includegraphics[angle=90,width=5in]{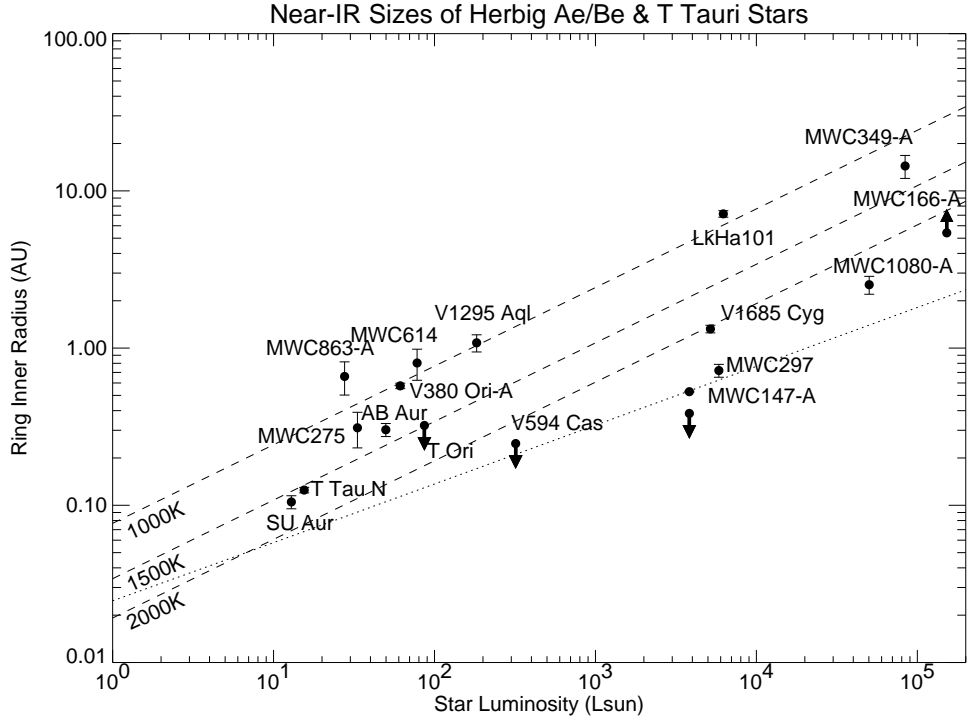}
\footnotesize
\caption{\footnotesize
This ``size-luminosity diagram'' is a compendium of all published
near-IR disk sizes of YSOs. The plot symbols show the radius of dust
emission for the ``ring'' model discussed in text.  The dashed lines
represent the expected inner edge of a dust disk truncated by dust
sublimation at temperatures $T_S=$1000\,K, 1500\,K, and 2000\,K,
assuming an optically-thin inner cavity and grain absorption
efficiency $Q_{\rm abs}=1$.  The dotted line shows the radius of dust
destruction (for $T_S=$1500\,K) predicted by the optically-thick,
spatially-thin disk model (the ``classical'' accretion disk)
previously held in favor (also $Q_{\rm abs}=1$).
\label{fig_theplot}}
\end{center}
\end{figure}

We begin by presenting Figure\,\ref{fig_theplot}, which contains in
one plot the complete set of published near-IR (1.6$\mu$m and/or
2.2$\mu$m) sizes of Herbig Ae/Be and T~Tauri disks (excluding the
accretion-dominated source FU~Ori).  
The extended ``disk'' component of each source was fitted with a 
ring model, further explained and justified below, and
the ring radius is reported for each target star.
The majority of measurements come
from the IOTA interferometer \citep{rmg1999,rmgphd,rmg2001}.  The
Palomar Testbed Interferometer contributed the T~Tauri observations as
well as one Herbig size \citep{akeson2000, akeson2002}.
Lastly, aperture masking at the Keck telescope was used for
LkH$\alpha$~101 and MWC~349 \citep{lkha2001,mwc2001}.  The stellar
properties (luminosity, effective temperature, distance, etc.)  were
taken from the original interferometer papers when possible, unless
revised spectral types were reported in 
\citet[][for MWC~614, V1295~Aql, MWC~296]{mora2001}.

Since single-baseline interferometers typically only measure
characteristic sizes and do not easily produce true images, we must
use some model for the brightness distribution when discussing a
``size.''   
The first image of the near-IR emission from a nearly 
face-on YSO
was recently published by \citet{lkha2001}, revealing a thin
asymmetric ring of emission
\citep[edge-on disks have been observed before, e.g.][]{koresko1998}. 
Partly based on this result, we have
chosen to represent the interferometer measurements in
Figure~\ref{fig_theplot} by simple ring models.  That is, the
visibility data for each source were fit assuming the dust emission
comes from a thin ring centered on the stellar point source (the
fraction of light coming from the star was typically estimated by
fitting the broad-band spectral energy distribution).

The use of a ring model can also be motivated theoretically.
Near-IR ($\simle$3$\mu$m) dust emission can only come from the
hottest dust, and thus only emerges from a narrow range of disk
annulii nearest to the star.  Further, once the dust opacity becomes
$\simge$1, the inner disk shields the outer disk causing a further
drop in dust temperature. 
While the near-IR ``bump''
\citep{hillenbrand1992} in the spectra of most target sources indicate
that the H-band and (especially) K-band fluxes are probably dominated by 
thermal emission of grains, some fraction of the emission must also be from
scattered stellar light.
Although scattered light from the inner edge of the disk will not
cause a significant change in the observed size,  
light scattered in a large-scale halo or upper layers of a flared disk
could cause the interferometric size to be larger than the disk inner
radius, adding a bias toward larger sizes in 
the size-luminosity diagram (see \S\ref{sec:other} for further discussion)

Different authors implemented the ring model fitting in slightly different
ways and we have used the inner ring radii published in the original
papers whenever possible \citep[for details, see][]{rmgphd}.  
This inhomogenity adds only small scatter to the
diagram ($\sim$10\%) and is insignificant compared to the
intrinsic scatter and other sources of uncertainty.
The errors bars reflect only the reported uncertainties in the 
ring radii, and 
do not include the typically larger uncertainties in the assumed
distances and stellar effective temperatures.  
Note that for early spectral types
the temperatures can be uncertain by thousands of
Kelvin \citep[e.g. MWC~297 has a 5 spectral
subtype uncertainty;][]{mora2001}. 
With improved interferometric data,
it will
become critical to accurately account for these effects and model each
source individually.

\subsection{Selection Effects}
This first generation of YSO disk measurements focused on bright
near-IR sources and suffered from some obvious
selection effects. 
The sources typically have unusually
large near-IR excesses from thermal emission of hot dust,
suggesting that these YSOs could have large disk masses and/or
accretion rates (MWC~166 is an exception). 
The large quantity of hot dust surrounding the sample stars
also implies that the inner disks have not yet been cleared of dust,
for example by forming planets
\citep[e.g. TW~Hya,][]{calvet2002}. Hence, this sample is not designed
to study disk evolution, and these disks are all presumably ``young.''
Lastly, the sample will contain few edge-on disks
\citep[MWC~349 is an exception,][]{mwc2001}, since spectra of such 
sources tend to peak more in the mid-IR
\citep[e.g.][]{lopez1995}.

Interestingly and fortunately, these selection effects do not strongly
influence our final conclusions, which ultimately depend most
sensitively on the dust grain properties and not the local
circumstellar environment.

\subsection{Basic Discussion of the ``Size-Luminosity Diagram''}
From inspection of Figure\,\ref{fig_theplot}, there is clearly a trend
of larger disk inner radii with increasing stellar luminosity
L$_\ast$, as one
would expect if the inner dust disk (or envelope) 
were truncated by dust evaporation
close to the star.  
  In Figure~\ref{fig_theplot} we have also plotted
the predicted location of the hottest dust as a function of
stellar luminosity for two simple models: an optically-thick,
spatially-thin disk model \citep[``classical'' accretion disk
after][assuming dust sublimation temperature
$T=1500$\,K]{hillenbrand1992} and a model with an optically-thin inner
cavity \citep[radius of dust sublimation $R_s\propto L_\ast^\frac{1}{2}$,
after][for three different assumed sublimation temperatures 
$T=$1000, 1500, 2000\,K]{lkha2001}.

``Classical'' accretion disks have a surface temperature $T\propto
r^{-\frac{3}{4}}$ since all the heating radiation hits the disk
obliquely \citep{friedjung1985}, the proportionality constant
depending on the size and temperature of the star.  In order to plot
predictions of models which are not functions purely of luminosity, we
must relate the stellar properties (e.g, effective temperature) to the
luminosity.  We have used here an empirical relation fitted to our
sample stars, $L_\ast \propto T_\ast^{5.3}$, similar to the slope of
the main sequence on the Hertzsprung-Russell diagram.  This allows us
to plot our size predictions purely as a function of luminosity; the
aggregate properties of our sample are sufficient for this Letter but
future detailed modeling must take into account the observed stellar
properties of each system individually.

As is evident in Figure~\ref{fig_theplot}, the classical accretion
disk model predictions of the near-IR sizes are too small for
nearly every target in our sample.  However, the larger sizes
predicted by the optically-thin cavity models seem to fit the
measurements better \citep[in agreement with the conclusions of][for
Lkh$\alpha$~101]{lkha2001}, although there is large amount of scatter
in the diagram for the Herbig Ae/Be stars, as noted by
\citet{rmg2001}.  These dust sublimation radii have been estimated as
described in detail below, neglecting backwarming by hot grains.  One
might think that uncertain distances could be a dominant source of
scatter in this diagram.  
 However, for a given source, a change in the
adopted distance will change both the physical ring inner radius and
the stellar luminosity in such a way as to move parallel to the
$R_s\propto L_\ast^\frac{1}{2}$ trend lines.

However, we have so far neglected to consider the optical properties
of the dust.  \citet{lkha2001} considered the case of dust sublimating
(at temperature $T_s$) due to heating only by the stellar radiation
(neglecting backwarming);
the dust sublimation radius $R_s$ follows the standard formula:
\begin{equation}
R_s = \frac{1}{2} \sqrt{Q_R} {\left( \frac{T_\ast}{T_s} \right) }^2 R_\ast
=1.1 \, \sqrt{Q_R} {\left(\frac{L_\ast}{1000\,L_\odot}\right) }^\frac{1}{2} 
{\left(\frac{T_s}{1500\,K}\right)}^{-2}~AU
\label{eq1}
\end{equation}
where $Q_R = Q_{\rm{abs}} (T_\ast) / Q_{\rm{abs}} (T_s)$ the ratio of the dust absorption
efficiencies $Q(T)$ for radiation at color temperature T of the
incident and reemitted field respectively.  

The dust destruction radius will be increased if one takes into
account backwarming by circumstellar dust, a msall correction 
dependent on the assumed dust distribution. We have also neglected any
effects of grains coupling to gas, which can cool through line
emission (this would {\em decrease} the dust destruction radius).  
However, we are not concerned with second-order
effects here and concentrate only on the dominant radiative 
processes in this Letter.

In the next section we consider how realistic dust properties 
dramatically affect the
predicted location of the dust destruction radius.

\section{Effects of Realistic Dust Properties}

Many workers have considered how the optical properties of cosmic dust
affect our interpretation of astronomical observations
\citep[e.g.,][and references therein]{dl84}. A particularly relevant set
of papers by \citet{wc1986, wc1987} considered spherically-infalling
dust onto a massive protostar.  Much of our argument parallels the
calculation of these authors (especially at the high-mass limit),
although the present application to T~Tauri and Herbig Ae/Be stars
span sufficiently different physical conditions to justify the
specific calculation presented below.

For the goals of this paper, it is
sufficient to consider the generic optical properties of
representative astronomical dust.  For the purpose of calculation we
have used the complex dielectric constants of \citet{ohm92} for warm
astronomical silicates and of \citet{dl84} for graphite.  Wavelength-
and grain size-dependent absorption efficiencies were calculated
assuming spherical grains and Mie scattering theory according to the
algorithm of \cite{toon81}.  Most of our discussion will focus on
silicate grains, and differences for carbon grains are noted at the
end of this section.

\begin{figure}[th]
\begin{center}
\includegraphics[width=5in]{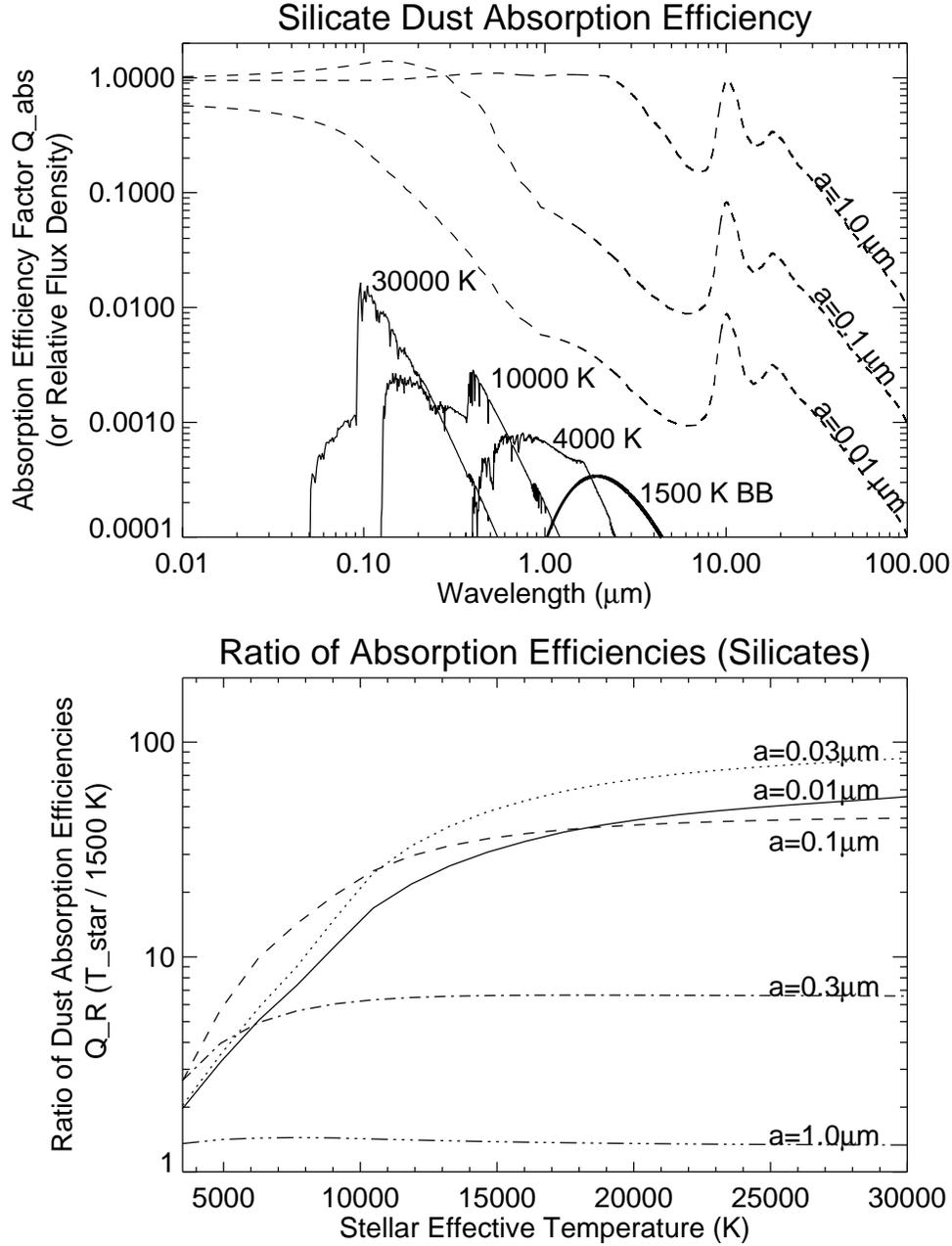}
\caption{
{\em top panel:} a) This figure shows the
wavelength-dependent absorption efficiency for silicate dust as a
function of dust grain radius $a$ (dashed lines).  At the bottom of the
plot, Kurucz models (log~g 4.1, log~Z 0.0) of stars at a range of
effective temperatures are plotted (solid lines) to illustrate why the effective
dust absorption efficiency is a strong function of spectral type for
small grains. {\em bottom panel:} b) This panel shows the ratio of the
dust absorption efficiencies for the incident radiation field versus
an assumed 1500\,K emission spectrum.  See text for mathematical
definitions.
\label{fig_qabs}}
\end{center}
\end{figure}

Figure~\ref{fig_qabs}a shows the absorption efficiency of silicate
grains of various sizes as a function of wavelength.  The
important property to note is that grains emit weakly at wavelengths
larger than the grain size.  Hence, small grains get heated to higher
temperatures for the same impinging radiation because they can not
cool efficiently.  Also included in this figure are Kurucz models
\citep[][and subsequent papers]{kurucz1979} for representative
spectral types spanning $1~L_\odot$ to $10^5~L_\odot$; note that
departures from a Planck function can be significant.

With these results, we can now calculate the factor $Q_R$ in
Eq.~\ref{eq1} for various dust sizes.  Formally,
the dust absorption efficiency 
for a given grain
size $a$ exposed to incident stellar radiation $I_\lambda(T_\ast)$, is defined as
follows:
$   Q_{\rm abs}(T_\ast) = \frac{\int Q_{\rm abs}(\lambda) I_\lambda(T_\ast) d\lambda }
		               {\int I_\lambda(T_\ast) d\lambda }$.
Figure~\ref{fig_qabs}b shows the ratio of absorption efficiencies
$Q_R$ as a function of stellar effective temperature, assuming an
emission temperature of 1500\,K (appropriate for dust near
sublimation).  For large silicate grains ($a\simge$1\,$\mu$m), $Q_R$
is relatively insensitive to the stellar effective temperature and
close to unity; this is simply because most of the radiation from
stars hotter than 3500\,K have wavelengths smaller than this grain
size.  However, the story is radically different for smaller grains.
For $a=0.1\mu$m, $Q_R$ values rise from a value of $\sim$8 for
$T_\ast=6000K$ to $\sim$50 for $T_\ast~\simge~20000K$.  Hence for a hot
star, reducing the dust size from 1~$\mu$m to 0.1~$\mu$m 
corresponds to 
nearly a factor of 50
increase in $Q_R$, resulting in
a factor of $\sim$7 increase in dust sublimation
radius (see Eq.\ref{eq1}).

\begin{figure}[t]
\begin{center}
\includegraphics[angle=90,width=6in]{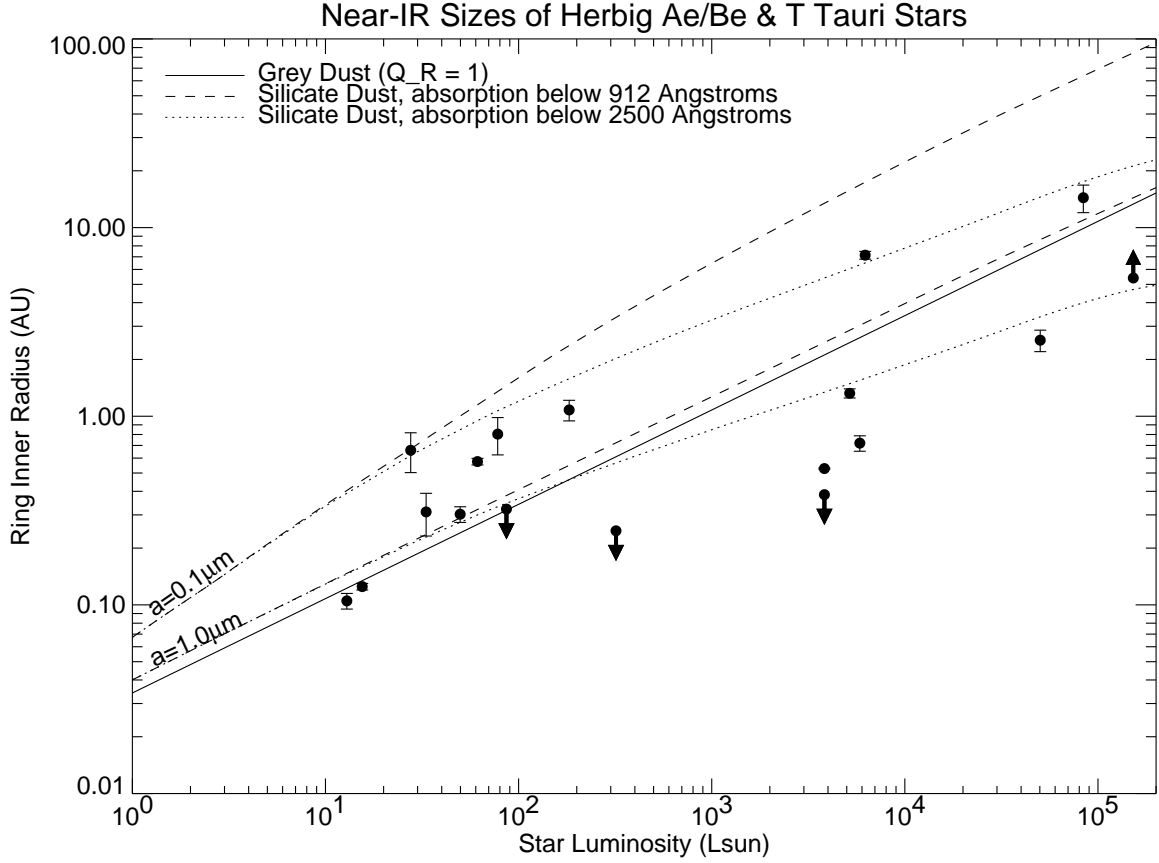}
\caption{
This ``size-luminosity diagram'' illustrates how 
the predicted disk sizes are modified by 
taking into account
realistic dust opacities and some gas absorption of the 
UV stellar continuum (all assuming $T_s=1500 K$).
The solid line shows radius of dust destruction for grey dust (see
$T_S=1500 K$ line in Figure\,\ref{fig_theplot}).
The dashed line assumes the Lyman continuum photons ($\lambda<912$\AA) 
have all been scattered out of the disk midplane, 
and the dotted line assumes
that the entire UV continuum below $\lambda<2500$\AA\ has 
been absorbed before impinging on the
dust (see \S\ref{section_gas} for discussion).  Curves for two different
grain sizes $a$ are shown.
\label{fig_theplot2}}
\end{center}
\end{figure}

Figure~\ref{fig_theplot2} shows how the dust sublimation radius (for
$T_S=1500K$) changes (compare to Figure~\ref{fig_theplot}) using these
results for silicate dust grain sizes $a$=0.1$\mu$m and 1.0$\mu$m.  In
creating these curves, we have also assumed that all stellar radiation
shortward of 912\,\AA\ was scattered away by hydrogen gas in the disk
midplane (see \S\ref{section_gas} for further discussion);
this slightly effects the predicted sublimation radius
for the only hottest stars.  

This figure shows clearly that dust grains with sizes similar to that
found in the interstellar medium \citep[0.01$\mu$m$ <a<
$0.25$\mu$m,][]{mrn} would not survive as close to the stars as is
observed, assuming $T_s \sim 1500K$.  However, larger grains with size
$a\simge$1.0$\mu$m are able to exist, thus reproducing the sizes
observed in Figure~\ref{fig_theplot}.  
This conclusion is more tentative for the T Tauris than for the Herbig stars, 
due to both the smaller sample size and the
less dramatic temperature difference between the photosphere and hot dust.
We note that large grains are expected
in the midplane of an accretion disk, given
that there are sufficient densities and timescales for significant
grain growth, even planet formation \citep[e.g.,][]{beckwith1991}.

\label{section_gr}
We have also performed these calculations for graphite dust and arrived
at the same basic conclusions.  As seen for silicates, it is the grain
size that has the strongest effect on $Q_R$ and the near-IR
sizes are most consistent with ``large'' carbon grains
($a\simge0.3\mu$m).  Note that the carbon grains can be somewhat
smaller (a factor of 3) than the silicate dust to reproduce the
characteristic sizes in Figure~\ref{fig_theplot2}, due to higher
near-IR absorption efficiency.

\section{Gas Absorption in the Inner Disk}
\label{section_gas}
In Figure~\ref{fig_theplot}, we see that disks around the higher
luminosity stars tend to be somewhat smaller in size than the lower
luminosity sources, compared to the $R_s \propto L_\ast^\frac{1}{2}$
trendline (see Eq.\,\ref{eq1}).  In the last section, we found
that realistic dust opacity effects make this discrepancy even greater
(see Figure~\ref{fig_theplot2}).

We suggest that a relatively low-density gas disk in the inner
cavity can significantly scatter the ultraviolet continuum from the
star, shielding the dust and hence shrinking the near-IR disk
sizes.  This effect will only occur for hot stars with significant
ultraviolet luminosity and could qualititively explain the trend
towards under-sized disks for the high luminosity YSOs.
\citet{wc1986} considered a similar effect in calculating the
equilibrium temperature structure around an accreting massive
protostar.   Indeed, \citet{hartmann1993} showed that the
gaseous inner disk of a Herbig Ae/Be star can become optically thick
for high accretion rates ($\dot{M}\simge 10^{-7}M_\odot$).  But what
about at the lower accretion rates currently favored 
\citep{hartmann1993,gullbring2000}?

Performing the proper calculation of this effect is beyond the
scope of this Letter, however we do consider one limiting case here.
We have already eliminated the Lyman continuum photons from the
impinging radiation field in creating Figure~\ref{fig_theplot2}, since
neutral H is so efficient at scattering such short wavelengths.  From
inspection of continuum opacity curves by \citet{novotny1973}, we see
that scattering from metals (Si, Fe, Mg) becomes an important opacity
source for $\sim$5000\,K gas for $\lambda\simle 2500$\AA.  At lower
temperatures, molecules can provide an additional source of opacity.
In Figure~\ref{fig_theplot2}, we have plotted additional curves for
the dust sublimation radius ($R_s$) assuming {\em all the energy}
shortward of $\lambda=2500$\AA\ has been removed from the radiation
field, a reasonable limiting case.  As expected, this can have strong
effects for the hottest stars, $L\simge1000\,L_\odot$, and should
motivate a quantitative study of this effect for realistic inner disk
gas models.  Ideally, modeling of this curvature in the
``size-luminosity'' relation will provide a unique and powerful
constraint on the density distribution of gas (total column density)
of the innermost (dust-free) disk.

We remark on an alternative interpretation of Figure~\ref{fig_theplot}
and Figure~\ref{fig_qabs}b. The near-IR disk sizes could be
explained by a classical accretion disk (optically-thick gas disk in
midplane) with {\em small} ($a\simle 0.1\mu$m) grains at the inner
radius, where the high $Q_R$ compensates for the loss of stellar 
energy absorbed/scattered
by material in the midplane.
However, since the inner gas disk is not expected to be
optically-thick for typical T~Tauris and Herbig Ae/Be stars, we have
not favored this model in this Letter.

\subsection{Other Important Effects}
\label{sec:other}
Considering the large scatter in the size-luminosity diagram for the
Herbig Ae/Be stars (factor of about 3-10 for a given stellar
luminosity), there must be additional effects contributing to the
observed disk sizes than the grain sizes and gas opacity.  Some effects
which may be influencing the observed sizes include: additional
sources of luminosity 
\citep[see][]{hillenbrand1992,hartmann1993}, photoevaporation 
\citep{hollenbach1994,mwc2001}, turbulent accretion \citep{kuchner2002a},
or more complicated emission 
morphology \citep[scattered light, viewing angles, ``disk plus envelope'' 
combinations, etc.;][]{mirosh1999}.

\section{Conclusions}
Physical models of accretion disks around young stars are critical for
understanding the level of IR excess, the spectral energy
denstribution, and the interferometric disk sizes.  However, such
models have a large number of parameters and are subject to many
uncertain assumptions.  In this Letter we considered only the problem
of understanding the observed characteristic near-IR sizes of
YSOs.
By limiting our focus to just the inner disk, 
we have motivated a minimalist framework founded on the 
argument that the characteristic size of a given source
is directly related to the radius of dust sublimation.

We investigated quantitatively how the radius of dust sublimation
depends on realistic optical constants and gas absorption of the
stellar ultraviolet continuum, and we found that
the magnitudes of these effects are
strong functions of stellar effective temperature and grain size.  
Our results showed
that the observed sizes of YSOs are consistent with the presence of an
optically-thin cavity surrounding the star, only if the
near-IR emission arises from relatively large dust grains
($a\simge 1\mu$m for Herbig Ae/Be stars, somewhat smaller grains
are allowed for
T~Tauris) heated to the sublimation temperature
($T_s \sim1500K$).
For the hotter stars (Herbig Be), gas
absorption/scattering of the stellar ultraviolet continuum is likely
to be non-neglible and may further help explain the observed
size-luminosity relations.

Despite success at reproducing the typical sizes, we can not easily
explain the large scatter in the size-luminosity diagram for the
Herbig Ae/Be stars using this minimalist framework.  
More complete disk surveys are beginning with the 
Keck Interferometer and the Very Large Telescope Interferometer
to explore these systems in more detail.
Further, interferometer arrays, such
as IOTA, COAST, and CHARA,  image the disk emission from YSOs to
determine whether the ring morphology (a central assumption here) is
common to these systems or unique to Lkh$\alpha$~101.

\acknowledgments
The authors would like to thank many colleagues for discussions and comments:
N. Calvet, L. Hartmann, P. Tuthill, J.P. Berger, M. Kuchner,  
J. Aufdenberg, C. D. Matzner, and A. Natta. 
This material is based upon work supported by NASA
under JPL Contract 1236050 issued through the Office
of Space Science.

\bibliographystyle{apj}
\bibliography{apj-jour,HerbigSizes,Thesis}

%% Generally speaking, only the figure captions, and not the figures
%% themselves, are included in electronic manuscript submissions.
%% Use \figcaption to format your figure captions. They should begin on a
%% new page.

\clearpage

%% No more than seven \figcaption commands are allowed per page,
%% so if you have more than seven captions, insert a \clearpage
%% after every seventh one.

%% There must be a \figcaption command for each legend. Key the text of the
%% legend and the optional \label in curly braces. If you wish, you may
%% include the name of the corresponding figure file in square brackets.
%% The label is for identification purposes only. It will not insert the
%% figures themselves into the document.
%% If you want to include your art in the paper, use \plotone.
%% Refer to the on-line documentation for details.

%\figcaption[sgi9259.eps]{This is the first figure and it uses sgi9259.eps as
%its EPS figure file. \label{fig1}}

%% Tables should be submitted one per page, so put a \clearpage before
%% each one.

%% Two options are available to the author for producing tables:  the
%% deluxetable environment provided by the AASTeX package or the LaTeX
%% table environment.  Use of deluxetable is preferred.
%%

%% Three table samples follow, two marked up in the deluxetable environment,
%% one marked up as a LaTeX table.

%% In this first example, note that the \tabletypesize{}
%% command has been used to reduce the font size of the table.
%% Note also that the \label command needs to be placed 
%% inside the \tablecaption.

\clearpage

\end{document}